# Spectrum Sensing Techniques For Cognitive Radio Networks

## Fatima Salahdine

STRS Lab, National Institute of Posts & Telecommunications, Rabat, Morocco
Electrical Engineering Department, University of North Dakota, Grand Forks, USA
E-mail: fatima.salahdine@gmail.com

Abstract: In this chapter, we present the state of the art of the spectrum sensing techniques for cognitive radio networks as well and their comparisons. The rest of the chapter is organized as below: Section I.1, Section I.2, and Section I.3 present the spectrum management problem and the cognitive radio cycle as well as the compressive sensing solution; Section II.1 describes the spectrum sensing model; Section II.2 presents the existing spectrum sensing techniques, including energy, autocorrelation, Euclidian distance, wavelet, and matched filter based sensing. Finally, a conclusion is given at the end of the chapter.

## I.1 Spectrum Management and Cognitive Radio

Wireless networks and information traffic have grown exponentially over the last decade, which has resulted in an excessive demand for the radio spectrum resources [1][2]. The radio spectrum is a limited resource controlled by regulations and the recognized authorities, such as the federal communications commission (FCC) in the US. The current radio spectrum allocation policy consists of assigning the channels to specific users with licenses for specific wireless technologies and services. Those licensed users have access to that spectrum portions to transmit/receive their data, while others are forbidden even when those spectrum portions is unoccupied [3]. Recent studies reported that the spectrum utilization ranges from 15% to 85% in the US under the fixed spectrum allocation (FSA) policy [4]. FCC measurements also show that some channels are heavily used while others are sparsely used as illustrated in Figure 1 [5].

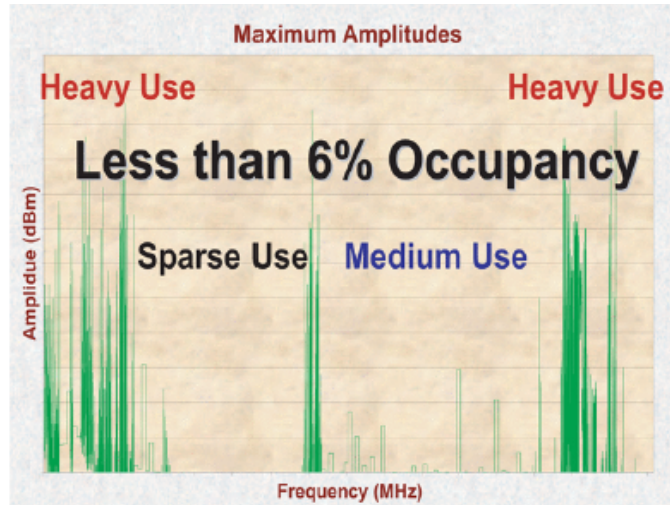

Figure 1: Radio spectrum occupancy [5][6][7].

Allocated spectrum portions are not used all the time by their owners, called primary users (PUs), which creates spectrum holes. A spectrum hole, also called white space, is a frequency band assigned to a PU, but it is not being used at a particular time and at a particular location. Therefore, the radio spectrum is inefficiently exploited [8][9]. Thus, the scarcity and inefficiency of the spectrum management require an urgent intervention to enhance the radio spectrum access and achieve high network performance. A better way to overcome the spectrum scarcity issue is dynamically managing it by sharing unoccupied channels with unlicensed users, called secondary users (SUs), without interfering with the PUs signals. The opportunistic spectrum access (OSA), also called dynamic spectrum access (DSA), has been proposed to address the spectrum allocation problems. In contrast to the FSA, DSA allows the spectrum to be shared between licensed and non-licensed users, in which the spectrum is divided into numerous bandwidths assigned to one or more dedicated users [10][11].

In order to advance the use of the OSA, several solutions have been proposed, including cognitive radio [12][13]. According to Mitola [14][15], cognitive radio is an intelligent radio frequency transmitter/receiver designed to detect the available channels and adjust its transmission parameters enabling more communications and improving radio operating behavior [14]. A cognitive radio system can observe and learn from its environment, adapt to the environmental conditions, and make decisions in order to efficiently use the radio spectrum. It allows SUs to use the PU assigned radio spectrum when it is temporally not being utilized as illustrated in Figure 2 [1][3].

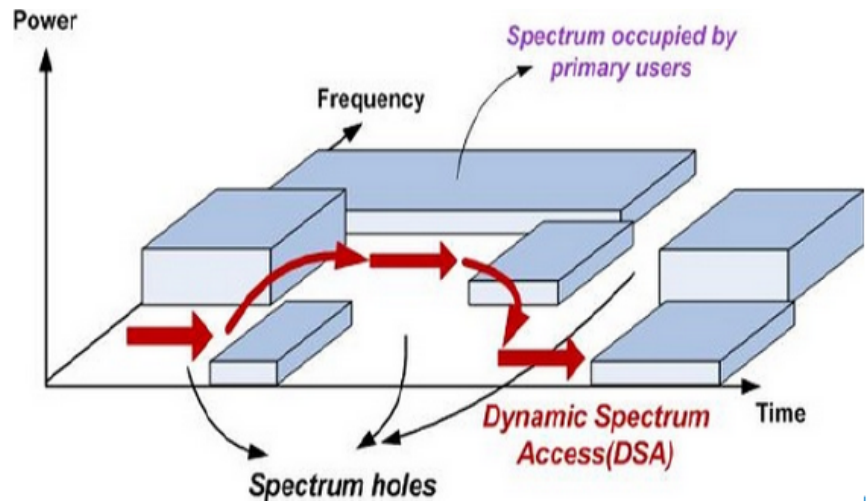

Figure 2: Dynamic spectrum access [3][16][17].

Cognitive radio is considered as the future technology to solve the resource allocation problem that the requirements of the 5th generation of the wireless communication raised. With these 5th generation of the wireless communication systems, the wide wireless will be interconnected offering high quality of service and data rates. The IEEE 802.22 standard has been defined as the first achievement of the cognitive radio solution to enable SUs to use the TV white spaces in the VHF and UHF bands [2].

## I.2 Cognitive Radio Cycle

As illustrated in Figure 3, a cognitive radio system performs a 3-process cycle: sensing, deciding, and acting [16][17].

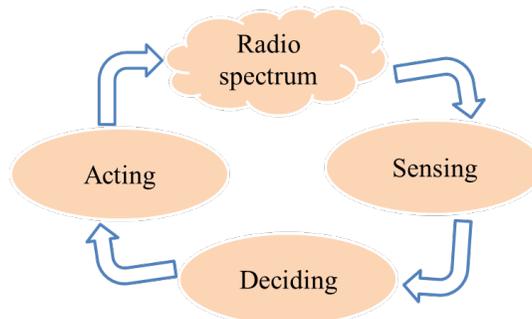

Figure 3: Cognitive radio cycle [17][16].

The first process is critical since it is the stage where the measurements are taken and the spectrum sensing is performed. Due to multipath fading, shadowing, or varying channel conditions [16][18]19], uncertainty affects this first process. In the observation process, measurements taken by the SUs are also uncertain. In the next process, SUs make a decision based on what has already been observed using their knowledge basis, which may have been impacted by the uncertainty in the detected measurements, leading to the wrong decisions. In the last process, uncertainty spreads over the cognitive radio cycle, and sometimes the wrong actions are taken [1]. Thus, uncertainty propagation impacts all the radio spectrum processes, which degrades the cognitive radio performance [16].

Therefore, it is necessary to address these uncertainty problems in the cognitive radio cycle by sensing the spectrum correctly, making the correct decision, and taking the right action.

## I.3 Compressive sensing

In order to sense the wideband radio spectrum, communication systems must use multiple RF frontends simultaneously, which can result in long processing time, high hardware cost, and computational complexity. To address these problems, fast and efficient spectrum sensing techniques are needed. Compressive sensing has been proposed as a low-cost solution for dynamic wideband spectrum sensing in cognitive radio networks to speed up the acquisition process and minimize the hardware cost [20]. It consists of directly acquiring a sparse signal in its compressed form that includes the maximum information using a minimum number of measurements and then recovering the original signal at the receiver. Over the last decade, a number of compressive sensing techniques have been proposed to enable scanning the wideband radio spectrum at or below the Nyquist rate. However, these techniques suffer from uncertainty due to random measurements, which degrades their performances. To enhance the compressive sensing efficiency, reduce the level of randomness, and handle uncertainty, signal sampling requires a fast, structured, and robust sampling matrix; and signal recovery requires an accurate and fast reconstruction algorithm [20-22].

Hence, efficient spectrum sensing and compressive sensing techniques are highly required in order to speed up the wideband spectrum scanning, deal with uncertainty, and perform accurate and reliable sensing occupancy measurements.

## II.1 Spectrum Sensing Model

Spectrum sensing is one of the most important processes performed by cognitive radio systems. It allows the SUs to learn about the radio environment by detecting the presence of the PU signals using one or multiple techniques and decide to transmit or not in its frequency band [1-5]. The spectrum sensing model can be formulated as:

$$y(n) = \begin{cases} w(n) & H_0: \text{PU absent} \\ h * s(n) + w(n), & H_1: \text{PU present} \end{cases} \quad (1)$$

where $n=1....N$, $N$ is the number of samples, $y(n)$ is the SU received signal, $s(n)$ is the PU signal, $w(n)$ is the additive white Gaussian noise (AWGN) with zero mean and variance $\delta_w^2$, and $h$ is the complex channel gain of the sensing channel. $H_0$ and $H_1$ denote respectively the absence and the presence of the PU signal. The PU signal detection is performed using one of the spectrum sensing techniques to decide between the two hypotheses $H_0$ and $H_1$. The detector output, also called the test statistic, is then compared to a threshold in order to make the sensing decision about the PU signal presence. The sensing decision is performed as:

$$\begin{cases} \text{if } T \geq \gamma, & H_1 \\ \text{if } T < \gamma, & H_0 \end{cases} \quad (2)$$

where $T$ denotes the test statistic of the detector and $\gamma$ denotes the sensing threshold. If the PU signal is absent, SU can access to the PU channel. Otherwise, it cannot access to that channel at that time. Figure 4 presents the general model of the spectrum sensing [23-26].

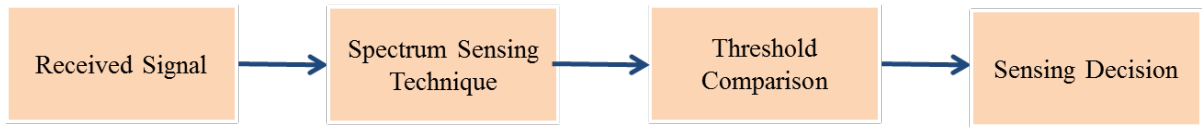

Figure 4: General model of spectrum sensing [26].

A number of sensing techniques have been proposed in the literature. These techniques are classified into two main categories: cooperative sensing and non-cooperative sensing as illustrated in Figure 5 [1][23].

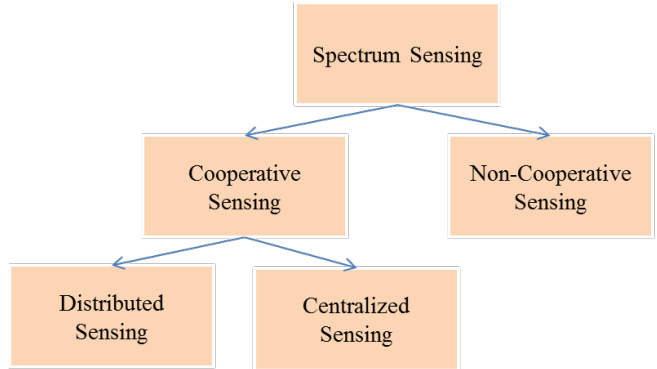

Figure 5: Spectrum sensing classification [24][25][26].

Under the non-cooperative sensing category, also called local sensing, each SU seeks for its own objectives and does not take into account the decisions of other SUs. As there is no communication or collaboration between the different SUs that sense the same frequency band, the spectrum sensing decision is performed locally. The non-cooperative techniques are simple and do not require high processing time and hardware cost. However, they are subject to errors due to shadowing, fading, interferences, and noise uncertainty. They are mainly adopted when only one sensing terminal is available or when there is no possible communication between the SUs.

Under the cooperative spectrum sensing category, the SUs collaborate and coordinate with each other taking into account the objectives of each user to make the final common decision. This cooperation between the different SUs can be divided into two schemes: centralized and distributed schemes. For the distributed scheme, SUs exchange their local observations and sensing results. Each SU takes its own decision taking into account the received results from the other SUs sensing the same frequency band. This approach does not require any common infrastructure for the final decision and the detection is controlled by the SUs. For the centralized scheme, all the SUs send their sensing results to a central unit, called fusion center, as illustrated in Figure 6.

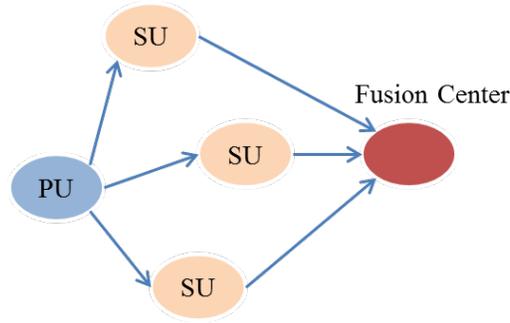

Figure 6: Centralized Cooperative spectrum sensing [24][25][26].

The fusion center decides about the spectrum access based on the received observations. The decision can be soft or hard combining decision with AND/OR rules. Under both spectrum sensing categories, SUs can perform the sensing using a spectrum sensing technique [23].

## II.2 Spectrum Sensing Techniques

A number of spectrum sensing techniques have been proposed to identify the presence of the PU signal transmission. These techniques provide more spectrum utilization opportunities to the SUs with no interferences or intrusive to the PUs. Examples of these techniques are presented in Figure 7, namely energy [1][24], autocorrelation [23], Euclidian distance [7][20], wavelet [4], and matched filter based sensing [6].

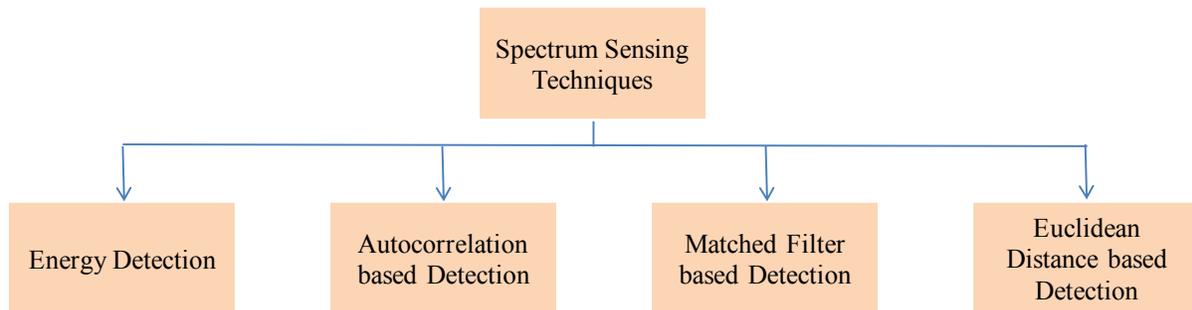

Figure 7: Examples of spectrum sensing techniques [24][25][26].

### II.2.1 Energy detection

Energy detection is the simplest sensing technique, which does not require any information about the PU signal to operate. It performs by comparing the received signal energy with a threshold. The threshold depends only on the noise power. The decision statistic of an energy detector can be calculated from the squared magnitude of the FFT averaged over *N* samples of the SU received signal as illustrated in Figure 8 [24][25-30].

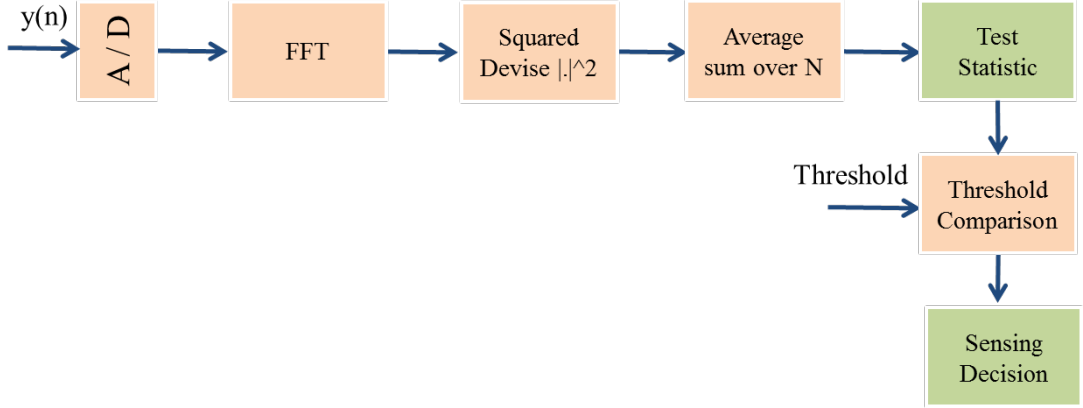

Figure 8: Energy detection model [25][26][30].

The detector output is the received signal energy as given by

$$T_{ED} = \sum_{n=0}^{N} y(n)^2 \qquad (3)$$

where *n=1…. N*, *N* is the sample number, and *y(n)* is the SU received signal, and $T_{ED}$ is the test statistic. Thus, the decision-based energy detection can be expressed as:

$$\begin{cases} \text{If } T_{ED} \geq \lambda, & \text{PU signal present} \\ \text{If } T_{ED} < \lambda, & \text{PU signal absent} \end{cases} \qquad (4)$$

where *λ* denotes the sensing threshold and $T_{ED}$ denotes the energy of the SU received signal. The received signal can be approximated as a Gaussian random signal. Based on the central limit theorem, when the number of samples exceeds 250 (*N* >250), the test statistic has a central chi-square distribution with *N* degrees of freedom for $H_0$ hypothesis, while it has a non-central chi-square distribution with *N* degrees of freedom for $H_1$ hypothesis. Thus, the test statistic, $T_{ED}$, is approximated as Gaussian and given by

$$\begin{cases} H_0: T_{ED} \sim \mathbb{N}(N\delta_w^2, 2N\delta_w^4) \\ H_1: T_{ED} \sim \mathbb{N}(N(\delta_w^2 + \delta_s^2), 2N(\delta_w^2 + \delta_s^2)^2) \end{cases} \qquad (5)$$

where $\delta_s^2$ denotes the PU signal variance, $\delta_w^2$ denotes the noise variance, and $\mathbb{N}$ denotes the normal distribution. For evaluation metrics, the probability of detection and the probability of false alarm for additive white Gaussian noise (AWGN) channel can be expressed respectively as:

$$Pd = Q\left(\frac{\lambda - N((\delta_w^2+\delta_s^2))}{\sqrt{2N((\delta_w^2+\delta_s^2))^2}}\right) \quad , \quad Pfd = Q\left(\frac{\lambda - N\delta_w^2}{\sqrt{2N\delta_w^4}}\right) \tag{6}$$

where $Q(.)$ denotes the Q-function and $\lambda$ denotes the sensing threshold. The two metrics can be also formulated as a function of the signal to noise ratio (*SNR*) as:

$$P_d = Q\left(\frac{\bar{\lambda} - N(1+\gamma)}{\sqrt{2N(1+\gamma)^2}}\right) \quad , \quad P_{fd} = Q\left(\frac{\lambda - N\delta_w^2}{\sqrt{2N\delta_w^4}}\right) \tag{7}$$

where $\gamma$ denotes the *SNR* and $\bar{\lambda}$ denotes the average threshold, $\bar{\lambda} = \lambda/\delta_w^2$. Thus, the sensing threshold depends on the noise power and it is expressed for a target $P_{fd}$ as:

$$\lambda = (Q^{-1}(P_{fd})\sqrt{2N} + N)\delta_w^2 \tag{8}$$

Each threshold value corresponds to a pair of ($P_d$, $P_{fd}$), representing what called the receiver operating curve (ROC). ROC represents the plotting of the correct detection rate as a function of the false detection rate for several thresholds [2][9][26][27][32–36].

Energy detection is easy to implement and does not require any prior knowledge about the PU signal, which makes it one of the most used techniques. However, it is very sensitive to the noise and cannot distinguish between the signal and the noise when the signal power is low. In addition, the sensing threshold for energy detector is an important parameter. When a detector does not adjust its threshold properly, it suffers from some performance degradation of the spectrum sensing. Various approaches were suggested for energy detection technique [31–37]. As the sensing performance is highly affected by the estimation error of the noise power, a dynamic estimation of the noise power is recommended in [31]. Adaptive threshold control is implemented with linear adaptation on the threshold based on the signal to interference plus noise ratio (SINR) [38][39]. This approach attains a considerably higher SU throughput than the fixed threshold approach, but maintains unacceptable chances of false alarms [32].

The authors in [33] presented an adaptive threshold in unknown white Gaussian noise with noise power estimation, keeping the false alarm rate at a preferred point under any noise level. This technique is based on a concept of dedicated noise estimation channel in which only noise is received by the SUs. An improved energy detection method is proposed in [34] where misdetection of the PU transmission due to a sudden drop in the PU transmission power is addressed by keeping an additional updated list of the latest fixed number of sensing events that are used to calculate an average test statistic value. A double-threshold technique is proposed in [35] with the intention of finding and localizing narrowband signals. Another technique is presented in [36] based on wideband spectrum sensing, which senses the signal strength levels within several frequency ranges to improve the opportunistic throughput of the SU and decreases the interference to the PU. In [37], the authors proposed an improved energy detection with an adaptive threshold to increase the detection rate.

## II.2.2 Autocorrelation Based Detection

Autocorrelation based sensing technique is based on the value of the autocorrelation coefficient of the received signal. It exploits the existing autocorrelation features in the transmitted signal and not in the noise [30]. For a given signal, *s(t)*, the autocorrelation function is defined as:

$$R_{s,s}(\tau) = \int_{-\infty}^{+\infty} s(t) * s^*(t-\tau)\, dt \qquad (9)$$

where $\tau$ denotes the time lag, *t* denotes time, and *s\** denotes the complex conjugate of the signal. In spectrum sensing context, sensing quality is affected by the noise level and it is difficult to interpret the signals affected by the Gaussian noise [40-44]. In fact, white noise is uncorrelated and its autocorrelation function results in a sharp spike at zero lag while the rest of lags are close to zero as illustrated in Figure 9.

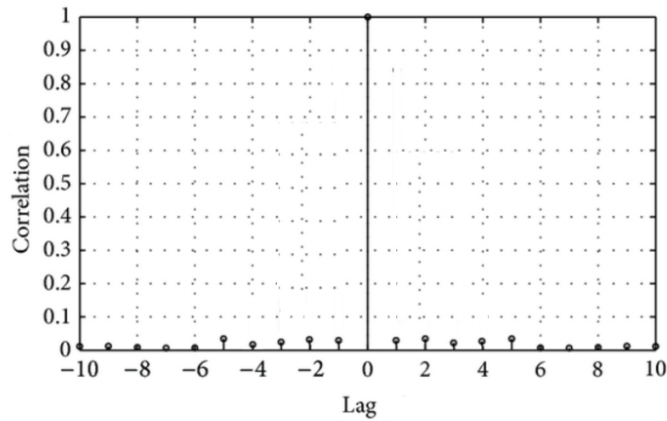

Figure 9: Autocorrelation function of the white noise [31][7].

However, $R_{s,s}(\tau)$ can present some high values depending on the transmitted stream proprieties. The transmitted signal is correlated; the zero lag and the first lag are very close as illustrated in Figure 10.

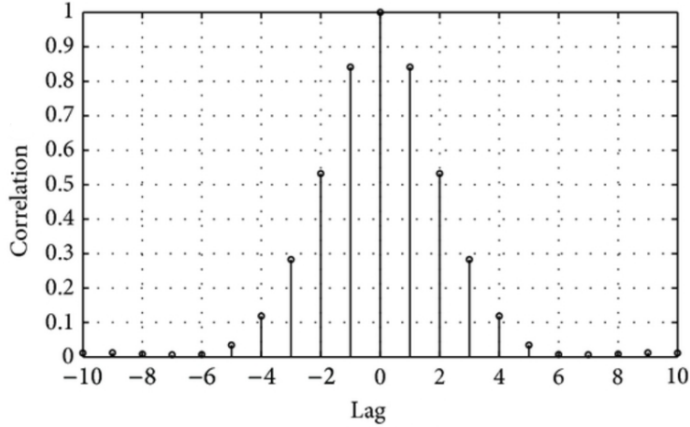

Figure 10: Autocorrelation function of the signal [31][7].

Therefore, the autocorrelation of the signal is correlated while the one of the noise in uncorrelated as illustrated in Figure 9 and Figure 10. When the degree of correlation is higher, the strength of the signal is higher [31][44]. Thus, the spectrum sensing is performed by exploiting the autocorrelation function to detect the PU signal presence under noise as illustrated in Figure 11.

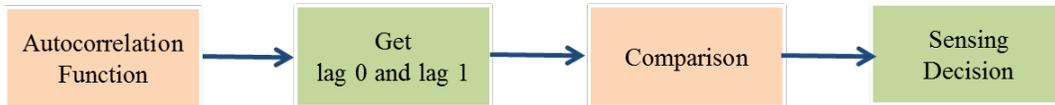

Figure 11: Autocorrelation based sensing model [31][7].

The sensing decision is based on the knowledge of the statistical distribution of the autocorrelation function. For random noise, the first lag of the autocorrelation is very small or negative, but when there is a signal the autocorrelation at the first lag represents a significant value. Thus, the sensing method consists on comparing *lag0* and *lag1* of the autocorrelation function of the SU received signal [41]. The sensing decision is expressed as:

$$\begin{cases} \text{if } lag0 \approx lag1 & \text{, PU signal present} \\ \text{if } lag0 \gg lag1 & \text{, PU signal absent} \end{cases} \quad (10)$$

The autocorrelation threshold is the margin between the two lag values. For instance, if *lag0* is superior to *lag1* by a value of $\lambda$ %, this value, $\lambda$, is the autocorrelation threshold. Autocorrelation based sensing is able to differentiate between signals and noise, which makes it less sensitive to noise uncertainty. It depends on the autocorrelation features and its performance is limited by the hardware based fractional frequency offset (FFO) for practical implementation; however, it is easy to implement and does not require high computing power [41].

## II.2.3 Euclidian Distance Based Detection

Euclidean distance based detection is a new sensing method proposed in [7]. It is mainly based on the autocorrelation of the SU received signal. This detector performs by computing the Euclidean distance between the autocorrelation of the signal and a reference line [7]. The autocorrelation of the received signal can be presented as the mean of

$$R_{s,s}(\tau) = \sum_{n=1}^{N} s(n) * s^*(n-\tau) \qquad (11)$$

where $R_{s,s}(\tau)$ is the autocorrelation at *lag$\tau$*, *s* is the received signal, and *N* is the number of samples. The reference line refers to the equation:

$$R = \left(\frac{2}{M}\right)t + 1 \qquad (12)$$

where *R* denotes the reference line, *M* denotes the number of lags of the autocorrelation including positive and negative values, and $0 \leq t \leq \frac{M}{2}$. The Euclidean distance, *D*, is the difference between the reference line and the signal autocorrelation [31]. It can be expressed as:

$$D = \sqrt{\sum (R_{s,s}(\tau) - R)^2} \qquad (13)$$

The sensing is then performed by comparing this metric with a threshold as illustrated in Figure 12.

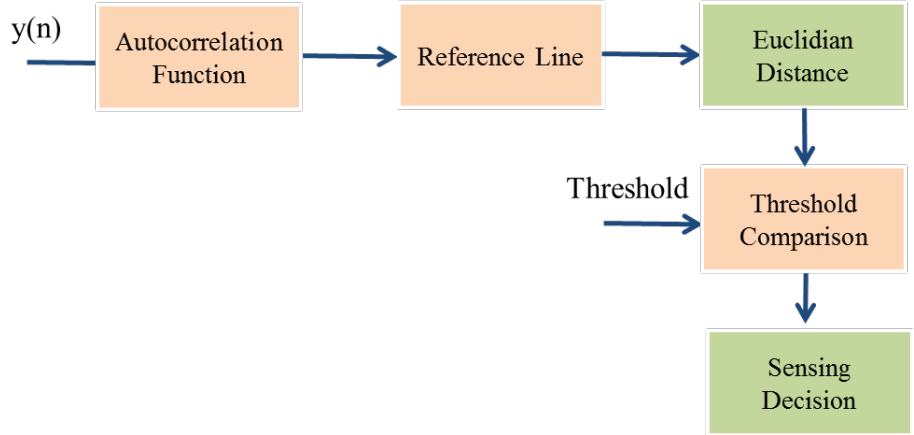

Figure 12: Euclidean distance based sensing model [31][7].

The sensing decision is expressed as:

$$\begin{cases} \text{If } D \geq \lambda, & \text{PU signal absent} \\ \text{If } D < \lambda, & \text{PU signal present} \end{cases} \tag{14}$$

where $\lambda$ denotes the sensing threshold. Euclidean distance based sensing is more efficient than the autocorrelation based sensing in terms of the detection success rate [40-44].

## II.2.4 Wavelet based Sensing

Wavelet based sensing, also called edge detection, is based on the continuous wavelet transform, which allows finding the signal decomposed coefficients with the help of a basis [41][42]. For a given signal $x(t)$, the continuous wavelet function, $\psi(t)$, operates in time domain where it includes a certain range and zeros elsewhere [45]. It is given by

$$f(s, u) = <x(t), \psi_{u,s}> = \int_{-\infty}^{+\infty} x(t) \psi_{u,s}^*(t) dt \tag{15}$$

where $s$ is the translating parameter, $u$ is the scaling parameter greater, and $\psi_{u,s}(t)$ is the basis. The wavelet transform allows mapping from the one dimensional signal, $x(t)$, to two dimensions coefficients $f(s,u)$. The frequency-time analysis can be performed at frequency corresponds to parameter, $s$, at time instant corresponds to parameter, $u$. The wavelet based sensing is operated by computing the continuous wavelet transform of the signal to perform the power spectral density. The local maximum of the power spectral density corresponds to the edge, which is compared to a threshold to decide about the spectrum occupancy as illustrated in Figure 13.

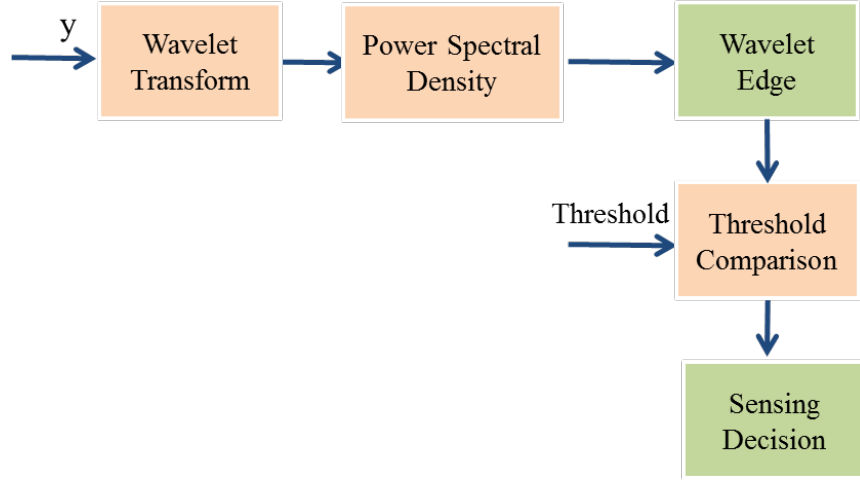

Figure 13: Wavelet based sensing system model [43][44].

The sensing decision is expressed as:

$$\begin{cases} \text{If } e \geq \lambda, & \text{PU signal absent} \\ \text{If } e < \lambda, & \text{PU signal present} \end{cases} \quad (16)$$

where $e$ denotes the wavelet edge and $\lambda$ denotes the sensing threshold. The wavelet edge is adopted for sensing decisions because the power density of a given signal corresponds to one spike at the signal frequency while it corresponds to multiple spikes when noise is added as illustrated in Figure 14.

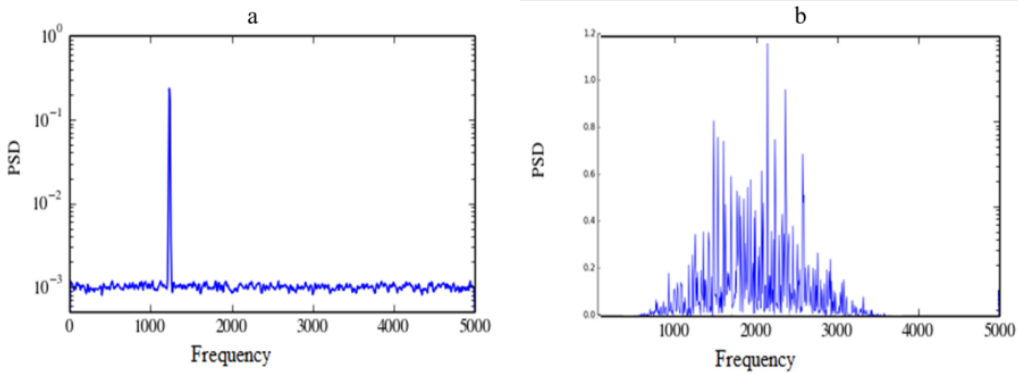

Figure 14: Power density function of (a) noiseless signal and (b) noisy signal [41][42].

It has been demonstrated that the wavelet based sensing can be adopted to identify the number of occupied bands over a frequency range, however, it requires high processing time [43]. Thus, wavelet based sensing can easily distinguish between the signal and the noise while deciding about the spectrum occupancy. It can also identify how many frequency bands are utilized. It represents less complexity but requires high processing time [45].

**II.2.5 Matched Filter Detection**

Matched filter detector is a coherent pilot sensor that maximizes the *SNR* at the output of the detector. It is an optimal filter that requires the prior knowledge of the PU signals. This sensing technique is the best choice when some information about the PU signal are available at the SU receiver. Assuming that the PU transmitter sends a pilot stream simultaneously with the data, the SU receives the signal and the pilot stream. Matched filter detection is performed by projecting the received signal in the direction of the pilot, *xp*, as illustrated in Figure 15 [6][30].

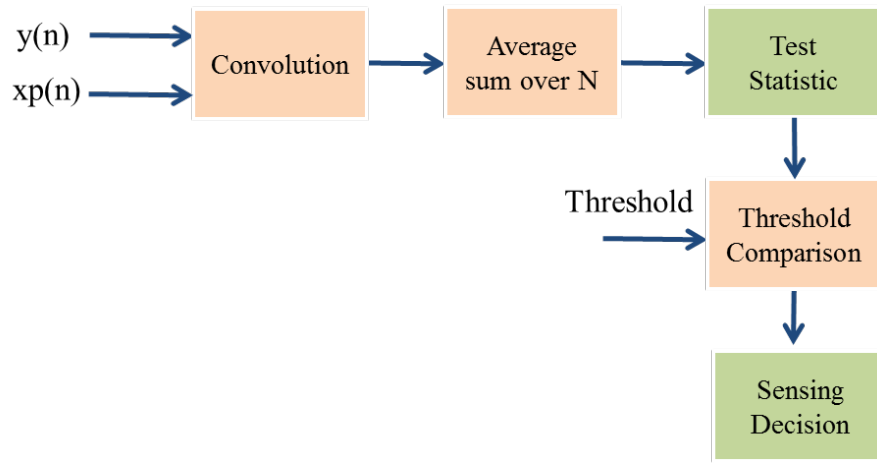

Figure 15: Matched filter detection model [30][7].

The test statistic is expressed as:

$$T_{MFD} = \sum_N y(n)\, x^*_p(n) \tag{17}$$

where $x_p$ denotes the PU signal, $y$ denotes the SU received signal, and $T_{MFD}$ denotes the test statistic of the matched filter detector. The test statistics, $T_{MFD}$, is then compared with a threshold in order to decide about the spectrum availability. The SU received signal, as well as the PU signal, are approximated to be random Gaussian variables. As a linear combination of Gaussian random variable, $T_{MFD}$ is also approximated as a Gaussian random variable.

$$\begin{cases} \text{If } T_{MFD} \geq \lambda, & \text{PU signal present} \\ \text{If } T_{MFD} < \lambda, & \text{PU signal absent} \end{cases} \tag{18}$$

Based on the Neyman-Pearson criteria, the probability of detection and the probability of false alarm are given by

$$Pd = Q(\frac{\lambda - E}{\sqrt{E\delta_w^2}}) \quad , \quad Pfd = Q(\frac{\lambda}{\sqrt{E\delta_w^2}}) \quad (19)$$

where $E$ is the PU signal energy, $\lambda$ is the sensing threshold, $Q(.)$ is the $Q$- function, and $\delta_w^2$ is the noise variance. The sensing threshold is expressed as a function of the PU signal energy and noise variance.

$$\lambda = (Q^{-1}(Pfd)\sqrt{E\,\delta_w^2} \quad (20)$$

Assuming that the signal is completely known is unreasonable and impractical. Some communication systems contain pilot stream or synchronization codes for channel estimation and frequency band sensing. A novel hybrid matched filter structure is proposed in [47], based on traditional matched filter by mixing segmented and parallel matched filter to overcome the frequency offset sensitivity. This new structure allows balancing between the sensing time and the hardware complexity. As both carrier frequency offset (CFC) and phase noise (NP) demean the sensing performance of matched filter detection, matched filter detection performance is examined in the presence of CFC and PN in [48]. Robust sensing technique is proposed to overcome the negative impact of CFC and NP on the performance and the ability of the sensing.

On the other hand, the sensing threshold for matched filter detector is an important parameter as for the other sensing techniques in which a number of researchers treated the threshold selection. In [49], the matched filter detection has been used with a static value to decide about the spectrum occupancy. In [50], each pair of ($P_d$, $P_f$) is associated with a particular threshold to make sensing decision. In other research works, the sensing threshold is determined dynamically by multiplying the theoretical threshold by a positive factor [18]. Others do not mention how the threshold was selected. Thus, with a static threshold, the sensing decision is not reliable because of the noise uncertainty. Therefore, the performance of the matched filter based detection depends mainly on the available information about the PU signal, including the bandwidth, central frequency, and modulation scheme. The sensing performance degrades when these data are incorrect or uncertain.

## II.3 Evaluation Metrics

To evaluate the performance of the spectrum sensing techniques, a number of metrics have been proposed, including the probability of detection, $P_d$, the probability of false alarm, $P_{fd}$, and the probability of miss detection, $P_{md}$. $P_d$ is the probability that the SU declares the presence of the PU signal when the spectrum is occupied [3][51][52]. The probability of detection is expressed as:

$$P_d = \text{Prob } (H_1/H_1) \tag{21}$$

where $H_0$ and $H_1$ denote respectively the absence and the presence of the PU signal. The higher the $P_d$, the better the PU protection is.

The probability of false alarm, $P_{fd}$, is the probability that the SU declares the presence of the PU signal when the spectrum is actually free (idle). It is expressed as:

$$P_{fd} = \text{Prob } (H_1/H_0) \tag{22}$$

The lower the $P_{fd}$, the more the spectrum access the SUs will obtain.

The probability of miss detection, $P_{md}$, is the probability that the SU declares the absence of a PU signal when the spectrum is occupied. It is given by

$$P_{md} = \text{Prob } (H_0/H_1) \tag{23}$$

These three metrics measure the efficiency of the spectrum sensing techniques and can be expressed as:

$$P_d + P_{fd} + P_{md} = 1 \tag{24}$$

There is a tradeoff between the probability of false alarm and the probability of miss detection. False detection of the PU activity causes interference to the PU and missed detection of the PU activity misses spectrum opportunities. This tradeoff can be expressed as conservative with $P_{fd}$ and aggressive with $P_{md}$; and a spectrum sensing technique has to fulfill the constraints on both probabilities [6].

## II.3 Conclusion

In this chapter, we presented the spectrum sensing and its categories. We described several spectrum sensing techniques and their characteristics, namely energy, autocorrelation, Euclidian distance, wavelet, and matched filter based detection. We also presented the different metrics used for performance evaluation.